\documentstyle[aps,epsf]{revtex}
\newcommand{\postscript}[2] {\setlength{\epsfxsize}{#2\hsize}
\centerline{\epsfbox{#1}}}
\begin{document}

\title{
Splitting of the $\pi - \rho$  spectrum
in a renormalized light-cone QCD-inspired model}

\author{T. Frederico}

\address{Dep.de F\'\i sica,
         Instituto Tecnol\'ogico de Aeron\'autica,
         Centro T\'ecnico Aeroespacial, \\
         12.228-900 S\~ao Jos\'e dos Campos, S\~ao Paulo, Brazil}

\author{Hans-Christian Pauli}
\address{Max-Planck Institut f\"ur Kernphysik, D-69029 Heidelberg, Germany}

\author{Shan-Gui Zhou}
\address{Max-Planck Institut f\"ur Kernphysik, D-69029 Heidelberg, Germany \\
         and \\
         School of Physics, Peking University, Beijing 100871, China}

\date{Draft \today}

\maketitle

\begin{abstract}
We show that the splitting between the light pseudo-scalar and
vector meson states is due to the strong short-range attraction in
the $^1S_0$ sector which  makes the pion and the kaon light
particles. We use a light-cone QCD-inspired  model of the mass
squared operator with harmonic confinement and a Dirac-delta
interaction. We apply a renormalization method to define the
model, in which the pseudo-scalar ground state mass fixes the
renormalized strength of the Dirac-delta interaction.\\
\\
PACS Number(s): 12.39.Ki, 11.10.Ef, 11.10.Gh, 14.40-n, 13.40.Gp
\end{abstract}

\pacs{12.39.Ki, 11.10.Ef, 11.10.Gh, 14.40-n, 13.40.Gp}

\section{Introduction}
\label{sec:intro}

The effective light-cone QCD theory \cite{pauli0,pauli2} is an
attempt to describe the Fock-state components of the light-front
meson wave-function of bound constituent quarks, constructed
recursively from the lowest Fock-state component. The lowest Fock
component of the hadron is an eigenfunction of an effective mass
squared operator parameterized in terms of an  interaction which
contains a Coulomb-like potential and a Dirac-delta term. Both
terms of the interaction come from an effective one-gluon-exchange
and the Dirac-delta corresponds to the hyperfine interaction. This
effective theory describes with reasonable success the masses of
the ground state of the pseudo-scalar mesons and in particular the
pion structure\cite{tob01}, inspite the small number of free
parameters, which is only the canonical number plus one --- the
renormalized strength of the Dirac-delta interaction. The model
does not have  confinement, thus the study of excited states above
the breakup threshold were not possible so far. The next step is
to introduce the confining interaction, therefore one should ask
as well, how the nice renormalization features of the model
changes and what are the implications of this model for the
spectrum of the pseudoscalar mesons.

In this work, we show an extension of the light-cone QCD-inspired
theory to include the confining interaction and we perform the
renormalization of the model, using as input the light
pseudo-scalar masses to fix the renormalized  strength of the
Dirac-delta interaction, active in the $^1S_0$-channel. We apply
the extended model to study the splitting of the excited
pseudo-scalar states from the excited $^3S_0$ vector meson states
as a function of the ground state pseudo-scalar mass. We show that
the $\pi - \rho$ mass splitting, due to the attractive Dirac-delta
interaction, is the source of the splitting between the masses of
the excited states, and inspite of the simplicity of the model,
there is a reasonable qualitative agreement with the experimental
values\cite{pdg}.

The non trivial mass splitting between the $\pi$ and $\rho$ meson
spectrum is a consequence of the small pion mass in the hadronic
scale, which fix the renormalized strength of the Dirac-delta
interaction. We must point out that the renormalized strength of
the Dirac-delta interaction, fitted to the pion mass, puts into
the model the  complex  short-range physics, which makes the pion
a strongly bound  system. In this first work, the Coulomb-like and
the confining interactions are substituted by an harmonic
oscillator potential, which allowed an analytic formulation, as we
are going to show. The parameters of the confining interaction in
the  mass squared operator, are fitted to the $^3S_1$ -meson
ground state mass and to the slope of the trajectory of excited
states with the principal quantum number\cite{anisov}. We expect
that the detailed form of the interaction using the Coulomb-like
potential and  a specific form of confinement will have a small
overall effect and will not change our conclusions about the
splitting of the pion and kaon spectrum in respect to the
correspondent vector meson spectra.

We should add that, the observation of the almost linear
relationship between the  mass squared of excited states with $n$,
as pointed out in Ref.\cite{anisov}, connects nicely and quite
naturally with our model. The slope of the trajectory in the $(n,
M^2)$ plane defines the harmonic oscillator strength of the mass
squared operator in our model. Here, we reveal some of the physics
that are brought by the  work of Ref.\cite{anisov} and we show the
relation between the $\pi$ and $\rho$ spectrum, through the pion
mass scale, which defines the renormalization condition of the
model.

This paper is organized as following. In section~\ref{sec:theory},
we review the light-cone QCD-inspired theory for the  mass squared
operator for a constituent quark-antiquark system and extend it to
include a confining interaction. We show how to renormalize the
theory using the subtracted equations for the transition matrix of
the model. In section~\ref{sec:pedestrian}, we present a
pedestrian approach to the theory developed in
section~\ref{sec:theory} using an harmonic oscillator interaction
in the mass squared operator equation, and show the results within
this model. In section~\ref{sec:remarks}, we made several remarks
on the qualitative aspects of the model  and we give the summary
of our work.

\section{Extended Light-cone QCD-inspired theory}
\label{sec:theory}

In this section we  extend the renormalized effective QCD-theory
given in Ref.\cite{tob01} to include confinement. The bare mass
operator equation for the lowest Light-Front Fock-state component
of a bound system of a constituent quark and antiquark of masses
$m_1$ and $m_2$ in the spin 0 channel, is described
as\cite{pauli0,pauli2}
\begin{eqnarray}
M^2\psi (x,{\vec k_\perp})&=&
\left[\frac{{\vec k_\perp}^2+m^2_1}{x}+\frac{{\vec k_\perp}^2+m^2_2}{1-x}
\right]\psi (x,{\vec k_\perp})
\nonumber \\
&-&\int \frac{dx' d{\vec k'_\perp}\theta(x')\theta (1-x')}
{\sqrt{x(1-x)x'(1-x')}}
\left(\frac{4m_1m_2}{3\pi^2}\frac{\alpha}{Q^2}-\lambda-W_{conf}(Q^2)\right)
\psi (x',{\vec k'_\perp}) \ ,
\label{p1}
\end{eqnarray}
where $M$ is the mass of the bound-state and $\psi$ is the
projection of the light-front wave-function in the quark-antiquark
Fock-state. The confining interaction is included in the model by
$W_{conf}(Q^2)$. The momentum transfer $Q$ is the space-part of
the four momentum transfer, the strength of the Coulomb-like
potential is $\alpha$ and $\lambda$ is the bare coupling constant
of the Dirac-delta interaction. For convenience the mass operator
equation is transformed to the instant form
representation\cite{saw1,pauli1}, with
\begin{eqnarray}
x(k_z)=\frac{(E_1+k_z)}{E_1+E_2} \ ,
\label{xkz}
\end{eqnarray}
and the Jacobian of the transformation of $(x,{\vec k_\perp})$
to $\vec k$ is:
\begin{eqnarray}
dx d\vec k_\perp= \frac{x(1-x)}{m_r A(k)}d\vec k \ ,
\end{eqnarray}
with the dimensionless function
\begin{eqnarray}
A(k)=\frac{1}{m_r}\frac{E_1 E_2}{E_1+E_2} \ ,
\label{phsp}
\end{eqnarray}
and the reduced mass $m_r=m_1m_2/(m_1+m_2)$. The individual
energies are $E_i=\sqrt{m_i^2+k^2}$ ($i$=1, 2) and $k\equiv|\vec
k|$.

The mass operator equation in instant form momentum variables is
given by:
\begin{eqnarray}
M^2\varphi(\vec k)&=&
\left[E_1+E_2\right]^2\varphi(\vec k)
\nonumber \\
&-&\int d\vec k'
\left(\frac{4m_s}{3\pi^2}\frac{\alpha}{Q^2\sqrt{ A(k)A(k')}}-
\frac{\lambda}{m_r\sqrt{ A(k)A(k')}}
+\frac{W_{conf}(Q^2)}{m_r\sqrt{ A(k)A(k')}}
\right)
\varphi (\vec k') \ ,
\label{mass1}
\end{eqnarray}
where $ m_s=m_1+m_2$, the phase-space factor is included in the
factor $1/\sqrt{ A(k)A(k')}$, $\sqrt{x(1-x)}\psi (x,\vec
k_\perp)=\sqrt{A(k)}\varphi(\vec k)$. The momentum transfer is
chosen in a rotationally invariant form as $Q^2=|\vec k
-\vec{k'}|^2$.

The bound state masses comes from the diagonalization of the mass
squared operator, however the mass operator of Eq.(\ref{mass1}) is
ill-defined mathematically. The bound states masses are, as well,
the poles of the Green's function of the theory. In view of the
necessary regularization and renormalization of the model due to
the singular interaction, we use the renormalization method of
singular interactions developed in the context of nonrelativistic
scattering equation\cite{t2} to derive the T-matrix and from which
we obtain the poles of the Green's function.

\subsection{Mass Operator and Green's Function}
\label{subsec:mass-operator}

The operator form of
Eq.({\ref{mass1}) is written as:
\begin{eqnarray}
 \left( M_0^2 + V + V^\delta +V_{conf} \right) | \varphi \rangle
 = M^2 |\varphi \rangle
 \ ,
 \label{mass2}
\end{eqnarray}
where the free mass operator, $M_0 \ (=E_1+E_2) $, is the sum of
the energies of quark 1 and 2, $V$  is the Coulomb-like potential,
$V^\delta$ is the short-range singular interaction and $V_{conf}$
gives the quark confinement. According to Eq.(\ref{mass1}) the
matrix elements of these operators are given by:
\begin{eqnarray}
 \langle \vec k | V | \vec{k'} \rangle
 &=& -\frac{4m_s}{3\pi^2}
     \frac{1}{\sqrt{ A(k)}} \frac{\alpha}{Q^2} \frac{1}{\sqrt{A(k')}}
 \ ,
 \label{mecoul}
\end{eqnarray}
\begin{eqnarray}
 \langle \vec k | V^\delta | \vec{k'} \rangle
 = \langle \vec k | \chi \rangle \frac{\lambda}{m_r} \langle \chi | \vec{k'} \rangle
 = \frac{1}{\sqrt{A(k)}} \frac{\lambda}{m_r} \frac{1}{\sqrt{A(k')}} .
 \label{mesing}
\end{eqnarray}
and
\begin{eqnarray}
 \langle \vec k | V_{conf} | \vec{k'} \rangle
 = \frac{1}{\sqrt{A(k)}} \frac{W_{conf}(Q^2)}{m_r} \frac{1}{\sqrt{A(k')}} .
 \label{meconf}
\end{eqnarray}
The phase-space factor $A(k)$ is defined by Eq.(\ref{phsp}), and
$Q^2$ is the  momentum transfer squared. For convenience of the
formal manipulations, the form-factor of the separable singular
interaction is introduced and defined by $\langle \vec k | \chi
\rangle = 1/\sqrt{A(k)}$.

Due to the confining interaction, the complete basis states used
to describe the renormalized theory are the eigenstates of the
mass squared operator without the singular interaction:
\begin{eqnarray}
 \left( M_0^2 + V +V_{conf} \right) |n \rangle
 = M^2_n |n \rangle
 \ ,
 \label{massconf}
\end{eqnarray}
where $n$ gives the quantum numbers of the eigenstate $|n
\rangle$, and $M^2_n$ is eigenvalue of the mass squared operator.

Our renormalization procedure is based on Green's function method,
so we introduce the free Green's function of the theory, which is
given by:
\begin{eqnarray}
G_0(M^2)=\sum_n \frac{|n \rangle \langle n|}{M^2-M^2_n} \ .
 \label{g0}
\end{eqnarray}
The poles of Eq.(\ref{g0}) give the eigenvalues of
Eq.(\ref{massconf}).

The complete Green's function of the theory includes the
Dirac-delta potential, and it is solution of the following
equation:
\begin{eqnarray}
G(M^2)= G_0(M^2)+G_0(M^2) V^\delta G(M^2)
 \ .
 \label{g}
\end{eqnarray}
The poles of Eq.(\ref{g}), on the real axis, are the eigenvalues
of the mass squared operator of the complete theory given by
Eq.(\ref{mass2}).

\subsection{Renormalized T-matrix}

In close analogy to scattering theory one can introduce the
transition-matrix from which the full Green's function of the
theory can be derived:
\begin{eqnarray}
G(M^2)= G_0(M^2)+G_0(M^2) T_{\cal R}(M^2) G_0(M^2)
 \ ,
 \label{gt}
\end{eqnarray}
where the matrix elements of the renormalized transition matrix,
$T_{\cal R}(M^2)$, satisfy the subtracted form of the
Lippman-Schwinger equation\cite{tob01,t2}:
\begin{eqnarray}
\langle n| T_{\cal R}(M^2)| n' \rangle
 &=&
 \langle n |T_{\cal R}(\mu^2)| n' \rangle
\nonumber \\
&+& \sum_{m}
\langle n |T_{\cal R}(\mu^2) |m \rangle
\left( \frac{1}{M^2-M^2_{m}}-
\frac{1}{\mu^2-M^2_{m}} \right)
\langle m| T_{\cal R}(M^2)| n' \rangle
\ .
 \label{tvren7}
\end{eqnarray}

The subtraction point $\mu^2$ is arbitrary, and it can move
without affecting the results of the theory, due to that physical
constraint $T_{\cal R}(\mu^2)$ changes according to the
Callan-Symanzik equation
\begin{eqnarray}
 \frac{d}{d \mu^2} \langle n| T_{\cal R}(\mu^2)| n' \rangle
 = -\sum_{m}
\langle n| T_{\cal R}(\mu^2)| m \rangle
\left(\frac{1}{\mu^2-M^2_{m}}\right)^2
\langle m| T_{\cal R}(\mu^2)| n' \rangle
\ .
 \label{tren7}
\end{eqnarray}

To solve Eq.(\ref{tvren7}), the T-matrix at the subtraction point
should be found, taking into account the operator structure of the
singular interaction, which is  the separable form of
Eq.(\ref{mesing}), we write that
\begin{eqnarray}
\langle n| T_{\cal R}(\mu^2)| n' \rangle =
\langle n| \chi \rangle
{\lambda}_{\cal R}(\mu^2)\langle \chi | n' \rangle
 \ ,
 \label{trenv5}
\end{eqnarray}
where $ \lambda_{\cal R}(\mu^2)$ is the renormalized strength of
the Dirac-delta interaction, which is fixed by the ground state of
the pseudo-scalar, $^1S_0$, meson.

The solution of the subtracted scattering Eq.(\ref{tvren7}) with
the T-matrix at the subtraction point defined by Eq.(\ref{trenv5})
is easily found, and it reads:
\begin{eqnarray}
\langle n| T_{\cal R}(M^2)| n' \rangle =
\langle n| \chi \rangle
t_{\cal R}(M^2)\langle \chi | n' \rangle
 \ ,
 \label{trenv20}
\end{eqnarray}
where the reduced matrix element is
\begin{eqnarray}
 t^{-1}_{\cal R}(M^2)
 = {\lambda}_{\cal R}^{-1}(\mu^2)+
\sum_{m}|\langle \chi | m \rangle|^2
\left(\frac{1}{\mu^2-M^2_{m}}-
\frac{1}{M^2-M^2_{m}} \right)
\ .
 \label{trenv21}
\end{eqnarray}
The divergence in the sum presented in Eq.(\ref{trenv21}), is
exactly cancelled by the difference between the free Green's
function appearing in that equation. The reduced matrix element,
$t_{\cal R}(M^2)$, is as well, the result of the change in the
renormalized coupling constant due to the dislocation of the
subtraction point from $\mu$ to $M$. The masses of the interacting
system are given by the zeros  of $t^{-1}_{\cal R}$ which defines
the zero angular momentum states of the bound quark-antiquark
systems.

The physical condition given by the ground-state of the pion or
the other pseudo-scalar mesons are the input to define the reduced
matrix element of Eq.(\ref{trenv21}). Choosing the subtraction
point at the mass of the ground state of the pseudo scalar meson,
for example at the pion mass, this implies that the renormalized
strength is zero and
\begin{eqnarray}
 t^{-1}_{\cal R}(M^2) =
\sum_{m}
|\langle \chi |m \rangle|^2
\left(\frac{1}{\mu^2-M^2_{m}}-
\frac{1}{M^2-M^2_{m}} \right)
\ ,
 \label{trenv22}
\end{eqnarray}
which has a zero at the mass $\mu$. The value of $\mu$,
as the physical input, can be changed to study  the splitting the
$^1S_0$ and $^3S_0$ spectrum in a continuous way, as we are going
to discuss in the next section.

\section{ Splitting of the spectrum in a Pedestrian Approach }
\label{sec:pedestrian}

Now to allow an analytic treatment of the renormalized theory,
while keeping its physical content, we use equal mass constituent
quarks and simplify  Eq.({\ref{mass1}) to the form:
\begin{eqnarray}
\left( M^2_{h.o.}+
g \delta (\vec r) \right) \varphi (\vec r)= M^2 \varphi (\vec r)
 \ ,
 \label{mass3}
\end{eqnarray}
where the bare strength of the Dirac-delta interaction is $g$, and
the mass squared operator for the harmonic potential is
\begin{eqnarray}
M^2_{h.o.}= 4( k^2+ m_q^2)+ \frac{1}{64} w^2 r^2 + a
 \ ,
 \label{mass4}
\end{eqnarray}
in units of $\hbar = c =1$.

The eigenvalue Eq.(\ref{massconf}) is given now by
\begin{eqnarray}
\left( 4(-\nabla^2+ m_q^2)+ \frac{1}{64} w^2 r^2 + a \right)
\Psi_n(\vec r)= M^2_n \Psi_n(\vec r)
 \ ,
 \label{mass5}
\end{eqnarray}
where $n$ is the principal quantum number of the s-wave excited
states and $\Psi_n(\vec r)$ is corresponding to the eigenstate of the
harmonic oscillator.

The observation of the almost linear relationship between the mass
squared of the $\rho$ excited states with $n$, as pointed out in
Ref.\cite{anisov}, connects quite nicely to our model. The slope
of the trajectory in the $(n, M^2)$ plane defines $w$ for each
system, and the ground state mass determines $a$. The constituent
quark mass, $m_q$, is included by the mass of the vector meson
ground-state. In this pedestrian approach, we are using that the
constituent masses of the up-down and strange quarks are
degenerated and due to that we will fit each case separately. The
eigenvalues of Eq.(\ref{mass5}) are given by $M^2_n=nw+M^2_{(v.g.s)}$,
where $M_{(v.g.s)}$ is the ground-state mass of
the $^3S_1$ meson. (In the work of Ref.\cite{anisov}, in the place
of our $n$ it is used $n-1$.)

The reduced T-matrix of our pedestrian approach turns to be
\begin{eqnarray}
 t^{-1}_{\cal R}(M^2) =
(2\pi)^3\sum_{n}
|\Psi_n(0)|^2
\left(\frac{1}{\mu^2-M^2_{n}}-
\frac{1}{M^2-M^2_{n}} \right)
\ ,
 \label{trenv23}
\end{eqnarray}
and the value of the s-wave eigenfunction at the origin, is given
by\cite{qm}:
\begin{eqnarray}
\Psi_n(0)=\alpha^{\frac32}\left[ \frac{2^{2-n}}{\sqrt{\pi}}
\frac{(2 n+1)!!}{n!}\right]^\frac12
\ ,
 \label{wf0}
\end{eqnarray}
where $\alpha^{-1}$ is the oscillator length.

The final form the reduced T-matrix of our pedestrian model is
 \begin{eqnarray}
 t^{-1}_{\cal R}(M^2) =
(2\pi\alpha)^3\sum_{n=0}^\infty
\frac{2^{2-n}}{\sqrt{\pi}}
\frac{(2 n+1)!!}{n!}
\left(\frac{1}{\mu^2- n w - M^2_{(v.g.s)}}-
\frac{1}{M^2- n w - M^2_{(v.g.s)}} \right)
\ .
 \label{trenv24}
\end{eqnarray}
The zeros of Eq.(\ref{trenv24}) gives the eingevalues of the the
mass squared operator of Eq.(\ref{mass3}).

The parameters of the present model are $w$,  the ground state
mass of the $^3S_1$ meson and the renormalized strength of the
Dirac-Delta interaction. We have here, the same number of
parameters of the original model\cite{pauli0,pauli2,tob01}, which
is the canonical number of 7 ($\alpha$ and 6 constituent quark
masses) plus one, the strength of the renormalized Dirac-delta
interaction. In the next, we show the numerical results for the
splitting of the $\pi - \rho$ and $K - K^*$ spectrum, due to the
Dirac-delta interaction acting in the pseudo-scalar $^1S_0$
channel.

In Figure 1, we present our results for the $\pi - \rho$ mass
splitting for the first six levels as a function of the
ground-state pseudo-scalar mass $\mu$, which interpolates from the
pion to the rho meson spectrum. The value of the parameter
$w$= 1.39 GeV$^2$ is taken from Ref.\cite{anisov}. The large splitting
which happens  in the ground state, is diminished in the excited
states, although the model attributes consistently smaller masses
for the $^1S_0$ states compared to the respective $^3S_1$ ones.
The $\rho(1700)$ is suggested to be a D-wave meson\cite{anisov},
and in the figure it does not fit well in our picture for $^3S_1$
meson resonance.

Although, it was pointed out \cite{anisov} that the kaon trajectory is
more uncertain, we have addressed here the splitting between the
$^1S_0$ and $^3S_1$ excited states, because our objective is to
verify the consistency of the model. The results for the kaon
states $^1S_0$ and $^3S_1$, $K$ and $K^*$, are presented in figure
2. These calculations confirms the pattern which we have observed
in figure 1: the large difference of the masses of the ground
states translates to a much smaller difference for the excited
states, and the masses of the $^1S_0$ states are consistently
below the correspondent $^3S_1$ meson masses. The oscillator
parameter for the kaon system is fitted to the difference of the
masses squared of the $K^*(1410)$ and $K^*(892)$, with value of
$w=$ 1.19 GeV$^2$.

\section{Remarks on the qualitative aspects of the model and Summary}
\label{sec:remarks}

The pseudo-scalar mass spectrum represents a true challenge to
theoretical modelling outside the physics described by just the
confining interaction\cite{god85}. The splitting between the $\pi
- \rho$ mass,  and the huge mass difference of the $\pi$ to its
first resonant state $\pi(1300)$, have to be  found from the
fitting of the potential parameters with the hyperfine
(short-range) interaction specially tuned to the pion mass. The
successful description of the splitting among the S-wave hyperfine
pairs $\pi - \rho$, $K - K^*$ and so on, depends crucially on the
structure of the short-range part of the hyperfine potential, as
concluded in the seminal work by Godfrey and Isgur\cite{god85}.
The work of Anisovich, Anisovich and Sarantsev \cite{anisov}
showing the trajectories of the mesons in the $(n,M^2)$ and
$(J,M^2)$ planes, putted a phenomenological order to the meson
spectrum, although the driven physics of this nice observation was
not clear. In regard to this last aspect, we believe that the
present work reveals some of the physics of the work in
Ref.\cite{anisov}.

In essence, the main physics of our model are
given by two ingredients: the dynamics described by a
mass squared operator and the Dirac-delta interaction, which
brings to the model the small pion mass.
We should point out that in the literature, the
attractive short-range interaction has been extensively addressed
in the study of the meson structure and spectrum. Below, we
will discuss very briefly some of these previous works, that are
more closely related to the present one in that particular aspect.

The Godfrey and Isgur model was used to construct light-cone
wave-functions for the pion, rho-meson and nucleon\cite{card}, and
it was pointed out that a high-momentum tail in the wave-function
above 1 GeV/c\cite{card} is due to the short-range attraction. The
impact on the structure of the pion and rho mesons were tested in
electroweak observables with some sucess\cite{card}. However, it
was pointed out\cite{ji} that the existing electroweak data was
not sensitive enough to allow a definite phenomenological
conclusion about the presence of hard-constituent quarks in the
hadron wave-function.

Another approach to the meson spectrum within a covariant quark
model \cite{riken}, using the framework of instantaneous
Bethe-Salpeter equation, also pointed out the importance of the
short-range attraction in the meson spectrum, due to the large
mass splittings in the pseudo scalar ($J = 0$) sector, which in that
work was attributed to the two-body point-like part of 't Hooft's
instant induced $U_A(1)$ symmetry breaking
interaction\cite{tooft}.

The structure of the pion light-cone wave function, has been
recently probed experimentally by diffractive dissociation of 500
GeV/c $\pi^-$ into dijets which provided a direct measurement of
the pion valence-quark momentum distribution for the first
time\cite{ashery}. The result give a stringent test of the
asymptotic wave-function\cite{bl}, which matches the high momentum
tail of the pion  wave-function\cite{tob01}. In that sense, this
experiment supports our model with the singular interaction it
contains.

From the previous discussion, where we just mentioned some
examples, becomes clear the necessity of a strong short-range
attractive interaction to built the pion, as seen in the modelling
of the light pseudo scalar meson spectrum and in the recent
measurement of the pion valence wave function. In that sense, we
believe that the Dirac-delta interaction parameterizes the complex
short-range physics of Quantum-Chromodynamics, which is brought to
the model by the small pion mass. The splitting of the pion and
kaon spectrum from the respective $^3S_1$-meson  counterparts, are
a direct consequence of the Dirac-delta interaction and its
renormalization, the structure of the  mass squared operator
equation with confinement, with all the relativistic effects
implied by it.

In summary, we extended the  renormalized light-cone QCD-inspired
effective theory to include confinement in  the mass squared
operator, which has a Dirac-delta interaction. We applied the
model to study the splitting of the $\pi - \rho$ and $K - K^*$
spectrum, using a pedestrian approach to it, with an harmonic
oscillator interaction. In this way, we derived an algebraic
equation, where the solutions give the  mass squared of the
$^1S_0$ states, as a function of the ground-state of the
pseudo-scalar meson. The model accommodates naturally the large
splitting of the ground state of the light $^1S_0$ and $^3S_1$ and
its suppression in the excited states, in qualitative agreement
with the experimental data. At the same time the model provides a
sound physical basis to understand the systematics of the
$q\overline q$-states in the $(n,M^2)$ plane and as well
in the $(J,M^2)$ plane. Therefore, we have
shown that the extension of the light-cone QCD-inspired model to
include confinement while keeping its simplicity and
renormalizability, provides a reasonable picture of
the spectrum of light mesons.

\acknowledgments

TF thanks the Max Planck Institute of Heidelberg and specially to
Prof. Hans-Christian Pauli for the warm hospitality during the
development of this work. TF also thanks Conselho Nacional de
Desenvolvimento Cient\'\i fico e Tecnol\'ogico (CNPq) and Funda\c
c\~ao de Amparo a Pesquisa do Estado de S\~ao Paulo (FAPESP) of
Brazil for partial financial support. SGZ is partly supported by
the Major State Basic Research Development Program of China Under
Contract Number G2000077407 and by the Center of Theoretical
Nuclear Physics, National Laboratory of Heavy Ion Accelerator at
Lanzhou, China.

\newpage

\begin{figure}
\postscript{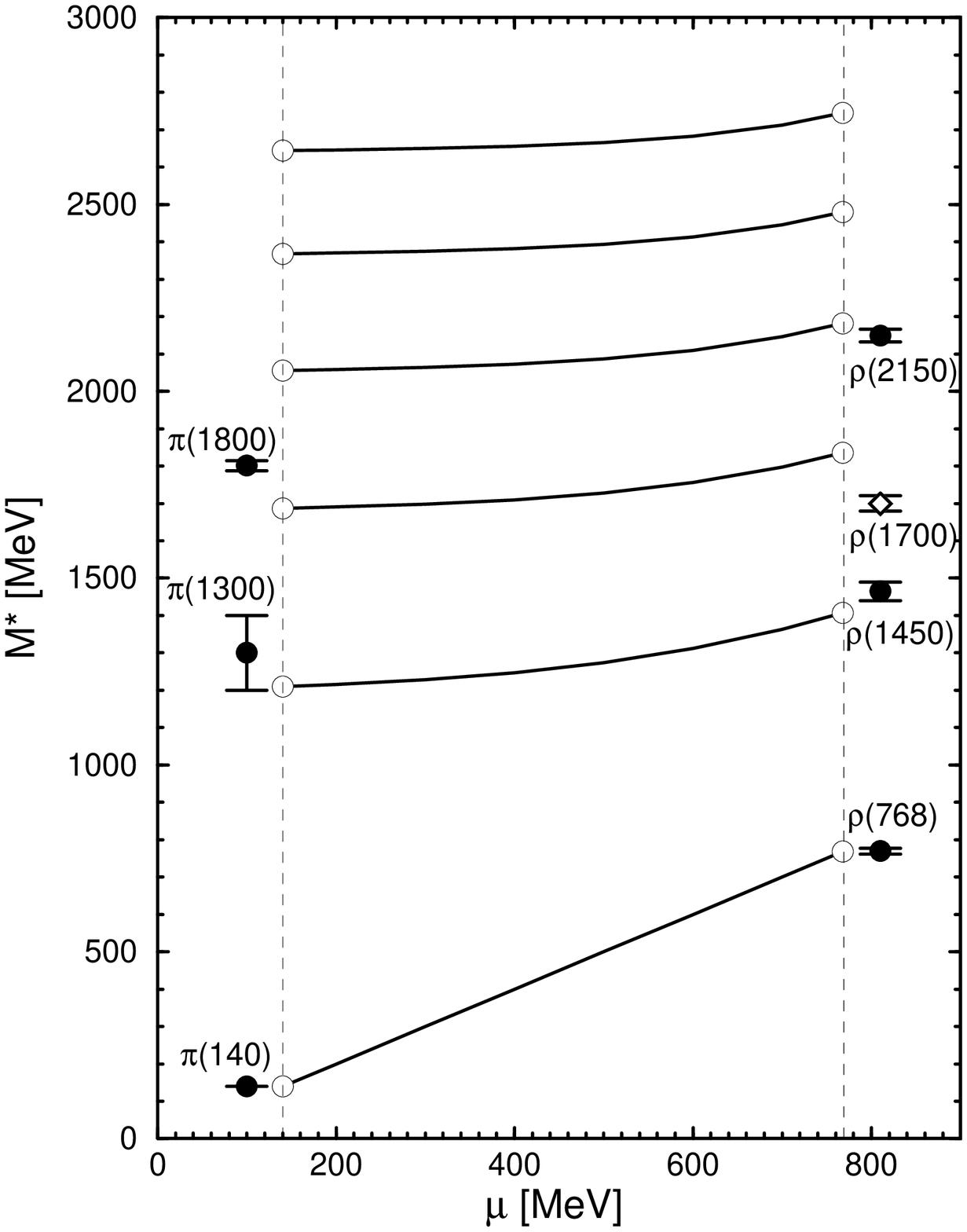}{.8}
\caption{Mass of the excited $q\overline q$ state ($M^*$) as a
function of the mass $(\mu)$ of the
pseudoscalar meson ground state for $I=1$.
Numerical results for $M^*$ from the
zeros of Eq.(\ref{trenv24}). The experimental data comes from
Ref.~\protect \cite{pdg}. The data on the left and right of the figure
corresponds to the $^1S_0$ and $^3S_1$ mesons, respectively.
}
\label{fig1}
\end{figure}

\begin{figure}
\postscript{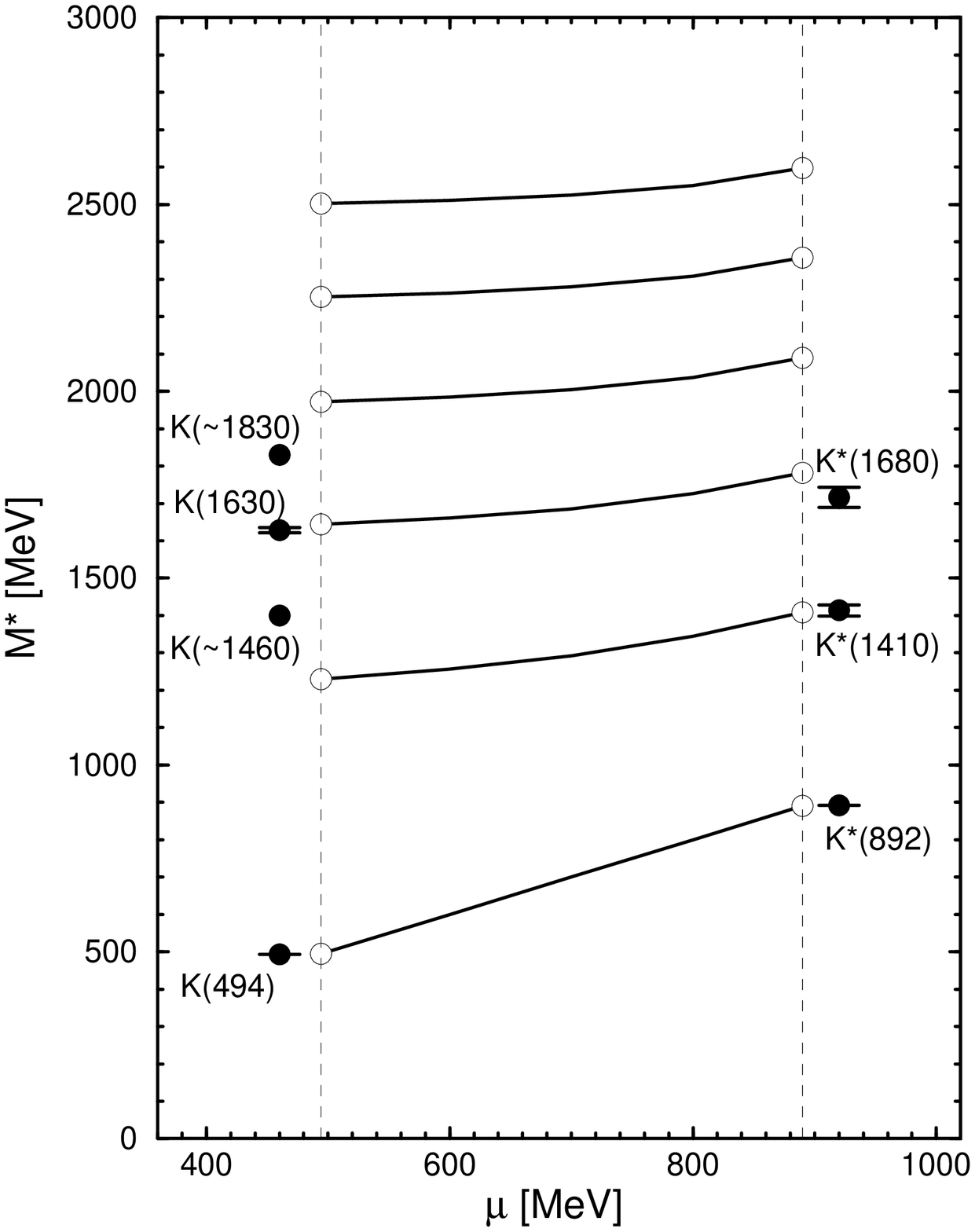}{.8}
\caption{
Mass of the excited $q\overline q$ state ($M^*$) as a
function of the mass $(\mu)$ of the
pseudoscalar meson ground state for $I=1/2$ of the strange sector.
Numerical results for $M^*$ from the
zeros of Eq.(\ref{trenv24}). The experimental data comes from
Ref.~\protect \cite{pdg}. The data on the left and right of the figure
corresponds to the $^1S_0$ and $^3S_1$ mesons, respectively.
}
\label{fig2}
\end{figure}

\end{document}